\documentclass[aps,prl,twocolumn,10pt]{revtex4-2}
\usepackage[T1]{fontenc}
\usepackage{amsmath,amsfonts,amssymb}
\usepackage{bm}
\usepackage{graphicx}
\usepackage{hyperref}
\usepackage{physics}
\usepackage{siunitx}
\usepackage[svgnames]{xcolor}
\usepackage{soul}
\usepackage{comment}

\newcommand{\beq}{\begin{equation}}
\newcommand{\eeq}{\end{equation}}
\newcommand{\bse}{\begin{subequations}}
\newcommand{\ese}{\end{subequations}}
\newcommand{\bea}{\begin{eqnarray}}
\newcommand{\eea}{\end{eqnarray}}
\newcommand{\bem}{\begin{displaymath}}
\newcommand{\eem}{\end{displaymath}}
\newcommand{\bmat}{\begin{bmatrix}}
\newcommand{\ebmat}{\end{bmatrix}}

\newcommand{\bc}{\begin{center}}
\newcommand{\ec}{\end{center}}
\newcommand{\bmk}{\bm{k}}
\newcommand{\bmr}{\bm{r}}

\begin{document}
	\title{Emergent fractals in hBN-encapsulated graphene based supermoir\'{e} structures and their experimental signatures}
	
	\author{Deepanshu Aggarwal}
	\affiliation{Department of Physics,
		Indian Institute of Technology Delhi,
		Hauz Khas, New Delhi 110016}
	
	\author{Rohit Narula}
	\affiliation{Department of Physics,
		Indian Institute of Technology Delhi,
		Hauz Khas, New Delhi 110016}
	
	\author{Sankalpa Ghosh}
	\affiliation{Department of Physics,
		Indian Institute of Technology Delhi,
		Hauz Khas, New Delhi 110016}
	
	
	\begin{abstract}
    Supermoir\'{e} structures (SMS), formed by overlapping moir\'{e}-patterns in van der Waals heterostructures, display complex behaviour that lacks a comprehensive low-energy theoretical description. We demonstrate that these structures can form emergent fractals under specific conditions and identify the parameter space where this occurs in hexagonal trilateral SMS. This fractality enables a reliable calculation of low-energy band counts, which are crucial for understanding both single-particle and correlation effects. Using an effective Hamiltonian that includes in- and out-of-plane lattice relaxation, we analyze SMS in hBN-encapsulated single and bilayer graphene. We prescribe methods to experimentally verify these fractals and extract their fractal dimension through angle-resolved photoemission spectroscopy (ARPES) and scanning tunneling microscopy (STM).
    \end{abstract}

	\maketitle

Twisted van der Waals (vdW) heterostructures \cite{Santos2007,Suarez2010,Santos2012} have exhibited intriguing single-particle physics \cite{Yankowitz2012, Wallbank2013} and strongly correlated electronic phases following the groundbreaking theoretical and experimental work on magic-angle twisted bilayer graphene (MATBLG) \cite{Bistritzer2010, Bistritzer2011,Cao2018I,Cao2018II}. While initial studies primarily focused on twisted bilayer systems \cite{Fang2020, Andrei2020, DAggarwal2023}, in recent years several intriguing experimental\cite{ZWang2019,Xu2021, Chen2021,XiZhang2021,Fischer2022,Zhang2022,Yiwei2022,Aviuri2023superconductivity, Devakul2023,Mjat2024,Kraig2024, Yazdani2024, Xia2025} and theoretical studies \cite{Morell2013, amorim2018, Bernevig2019, Kaxiras2020,Dumitru2021,Guinea2021,Maine2022,Popov2023, Qin2023, Jiabin2023, Nakatsuji2023, Meng2023, Mao2023,Imran2023,Parameswaran2024,Foo2024,Xia2025} were carried out in twisted multilayers (layer count $N >2$) indicating more complex behaviour with increasing $N$. 

In such multilayer vdW heterostructures, there may be a rotational misalignment (twist) between successive layers \cite{Geim2013} made of the same or different materials. Barring the lattice mismatch in the case of different materials, the most general case can be considered as the stacking of $N-1$ moir\'{e}-patterns arising out of $N$ twist angles over one another with varying length scales, thus resulting in a supermoir\'{e} structure (SMS) \cite{ZWang2019}. Particularly interesting is the $N=3$ case, where two moir\'{e}-patterns are stacked over one another. The characteristic length scale of such trilayers is the supermoir\'{e} (SM) wavelength formed by the interference of two periodic moir\'{e}-patterns of different interfaces, \textit{e.g.} twisted trilayer graphene \cite{Morell2013,amorim2018,Bernevig2019,Kaxiras2020,Qin2021,Guinea2021,Parameswaran2024}, hexagonal boron-nitride (hBN) encapsulated single-layer graphene (hBN-G-hBN) \cite{Lujun2019, Wang2019, Viti2020, Andelkovic2020, Koshino2021, Crosse2021, Miao2022, Guarochico2023} or hBN-encapsulated Bernal stacked bilayer graphene (hBN-BLG-hBN) \cite{Spethmann2021,Smeyers2023,Zollner2023,Shilov2024,Kareekunnan2024,Mjat2024} etc. which obey the condition: SM wavelength~$\ge$ moir\'{e} wavelength of individual interfaces. By considering the bandstructures of two generic trilayer SMS: hBN-G-hBN and hBN-BLG-hBN (a quad-layer system but still works effectively as a trilayer SMS), which include the effect of relaxation \cite{Carr2018,Carr2019}, we demonstrate the emergence of fractal behaviour in a limited region of phase space created by the relative twist of the two underlying moir\'{e}-patterns. We show how the fractal dimension $D_{f}$ accurately determines certain features of the bandstructure, thus profoundly affecting single-particle and correlation physics. We subsequently establish how experiments such as nano-ARPES and scanning probe techniques \cite{Angela2024} can identify such emergent fractality by explicitly demonstrating the signature of newly inserted bands due to the formation of the supermini zone in the reciprocal-space (RS) and experimentally determining $D_f$.

\begin{figure*}
	\centering
	\includegraphics[width=\linewidth]{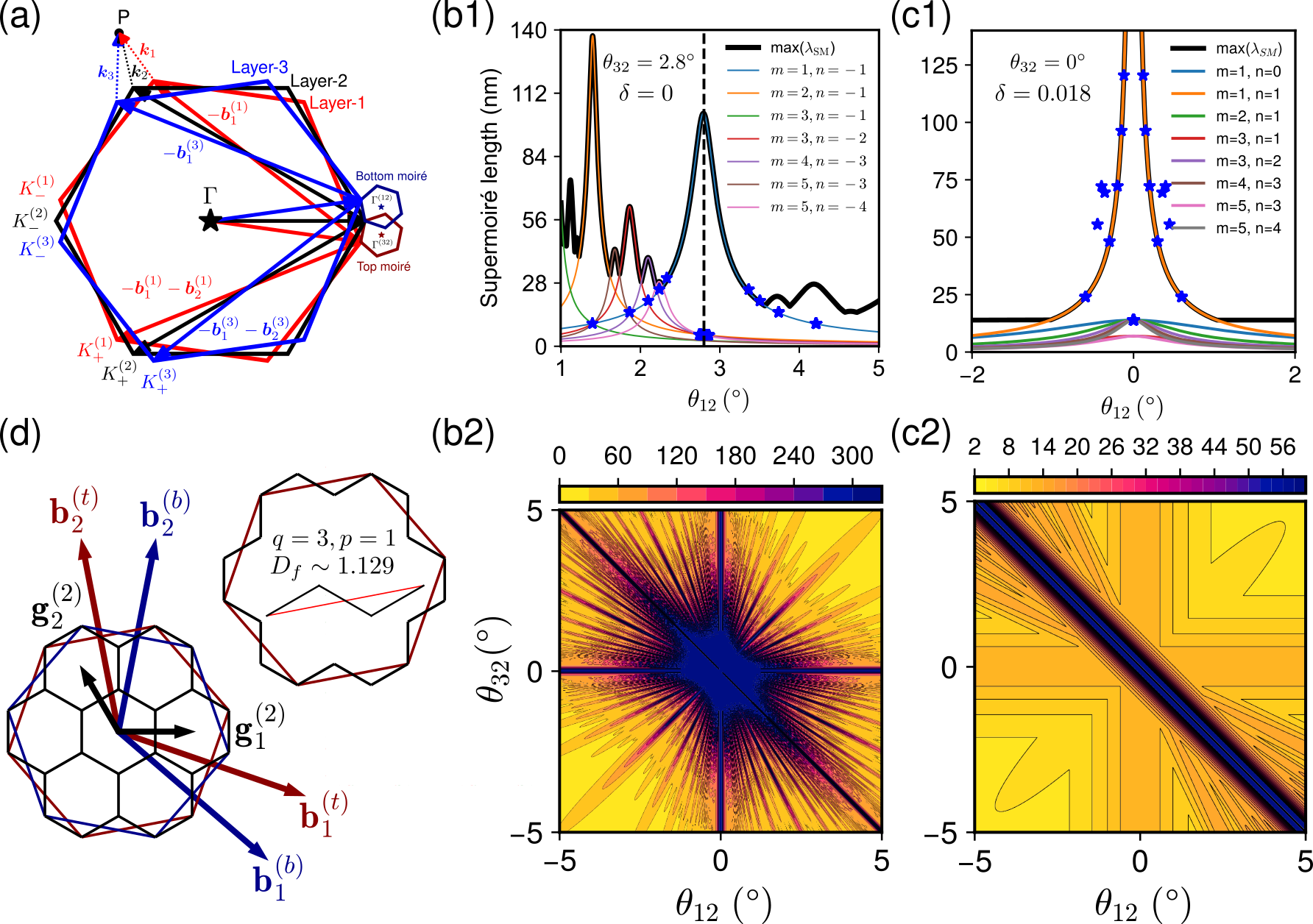}
	\caption{
(a) The BZ of the three individual layers superposed over one another, followed by their respective rotations \emph{wrt} the middle layer. $K^{(\ell)}_{-}$ marks the left valley at the BZ corners of layer-$\ell$ and $K^{(\ell)}_{+}$ marks the right valley of layer-$\ell$. The same flavour valleys are connected through the reciprocal lattice vectors such that the two vectors $\bm{b}^{(1)}_{1} - \bm{b}^{(1)}_{2}$ and $\bm{b}^{(3)}_{1} - \bm{b}^{(3)}_{2}$ connect the right valley of layer-1 and layer-3. The black star at the centre marks the high-symmetry $\Gamma$-point, while $P$ marks a generic point in the RS that can be reached from the $\Gamma$-point via the three nearest lying $K$-valleys plus the three vectors $\bmk_{1}$, $\bmk_{2}$ and $\bmk_{3}$ of each layer. The two small hexagons on the right identify the moir\'{e} BZ corresponding to two moir\'{e} interfaces between layer-1/2 and layer-3/2, which are also misoriented \emph{wrt} each other. (b1) The variation of the second-order moir\'{e} length is plotted against $\theta_{12} \in [\ang{1},\ang{5}]$ for a fixed angle $\theta_{32} = \ang{2.8}$ and the blue stars mark the various possible commensurations of around $\ang{2.8}$ (see text). (b2) The maximum second-order moir\'{e} length scale of two moir\'{e} interfaces is plotted by considering the range of angles $\theta_{12} = -\theta_{32} \in \big[\ang{-5},\ang{5}\big]$. This pattern is diagonally-symmetric about the diagonal $\theta_{23}=-\theta_{12}$. (c1), (c2) The same geometrical analysis is performed with a non-vanishing lattice mismatch $\delta = 0.018$, particularly depicting the hBN-encapsulated single-layer and bilayer configurations. (d) In this work, we are interested in the part of the parameter space where the moir\'{e}-patterns in two-layer interfaces make a commensurate angle with each other. Therefore, this geometry shows two moir\'{e} BZ (maroon and dark blue) of equal side length and the commensurate mini BZ (black) at $(q,p) = (3,1)$. The outer boundary of the mini BZ overlapping either of the mBZ provides a shape that can be generated using a generator \cite{Arlinghaus1985,Duvall1992,Aggarwal2024}.}
	\label{fig:systemtri}
\end{figure*}


Following \cite{Carr2018, Cazeaux2020}, the local stackings in the configuration and RS of a generic moir\'{e}-pattern can be written as a $ 2\times 2 $ matrix 
\begin{equation}
	A_{\ell} = \bmqty{\bm{a}^{(\ell)}_{1} & \bm{a}^{(\ell)}_{2}} = \mathcal{L}_{\ell}A;~G_{\ell} = 2\pi A^{-1,T}_{\ell}= \bmqty{\bm{b}^{(\ell)}_{1} & \bm{b}^{(\ell)}_{2}}
\end{equation}
Here $A = \bmqty{\bm{a}_{1} & \bm{a}_{2}}$ is a common matrix where the primitive vectors (PV) of both layers, $ \bm{a}^{(\ell)}_{i} = \bmqty{a^{(\ell)}_{ix} & a^{(\ell)}_{iy}}^{T} $ are the direct-space PV for the two periodic layers that create the moir\'{e}-pattern with $i=1,2$, and $ \ell=1,2 $ being the layer-indices with the application of an invertible linear mapping $ \mathcal{L}_{\ell} $ of $2\times2$ dimension (see Table-I, SI \cite{supp}), and $\bm{a}^{(\ell)}_{i} \cdot \bm{b}^{(\ell)}_{j} = 2\pi\delta_{ij}$. The superscript $T$ indicates the matrix transpose. The above definitions in the configuration and RS can also be applied to general SMS. For trilateral SMS (see the RS and BZ of a prototype homo-trilayer supermoir\'{e} in Fig. \ref{fig:systemtri} (a)), the reciprocal lattice vectors consisting of two moir\'{e} interfaces are
\begin{equation}
    \bm{b}^{\text{M}(\ell' \ell)}_{i} = \bm{b}^{(\ell)}_{i} - \bm{b}^{(\ell')}_{i} = \left(\mathcal{I} -\mathcal{L}_{\ell'}\mathcal{L}^{-1}_{\ell}\right)\bm{b}^{(\ell)}_{i} \quad \forall \quad i=1,2,
    \label{eqn:moir\'{e}recivecs}
\end{equation}
defining SM harmonics labelled by the integer pair $(m,n)$, such that the primitive reciprocal lattice vectors are given as the columns of the matrix $G_{\text{SM}}(m,n)$ (Eq.4, Sec-I SI \cite{supp}). However, the SM length scales derived from these harmonics (Eq.5a to Eq.5i and the related discussion in SI \cite{supp}) do not provide the representative periodicity in SM structures in such twisted trilateral SMS as for MATBLG, making the construction of an effective low-energy theory a challenge \cite{Kaxiras2020,Koshino2022}. Considering hexagonal supermoir\'{e} systems such as graphene layers or graphene layers encapsulated by hBN, we determine that at certain twist angles, such SMS are emergent fractals which provide insight about their low-energy bandstructure and correlate these findings with recent experimental techniques developed to probe such SM structures.

The commensuration between the two moir\'{e} interfaces results in a supermoir\'{e} commensurate cell which implies the existence of a linear mapping $\mathcal{L}=A^{-1}\mathcal{L}^{-1}_{12}\mathcal{L}_{32}A$ for the two moir\'{e} interfaces such that it maps a pair of integers $(n'_{1}, n'_{2})$ to $(n_{1}, n_{2})$, where $\mathcal{L}_{\ell 2} = \big[\mathcal{I} - \mathcal{R}^{-1}(\theta_{\ell 2})\big(1+\delta_{\ell 2}\big)^{-1}\big]^{-1}$. For a solution to exist, each matrix element of $\mathcal{L}$ must be a rational number. The blue stars in Fig. \ref{fig:systemtri}(b) and (c) correspond to approximate commensurate structures that are numerically obtained within a tolerance value of 0.1, giving points in the parameter space of $\theta_{12}$ and $\theta_{32}$ where an approximate commensurate supermoir\'{e} cell exists.

Of particular interest, when the moir\'{e} length scales in the two interfaces are equal, $\lambda^{(t)}_{12} = \lambda^{(b)}_{23} = \lambda_{0} $, \emph{i.e.}, the two moir\'{e}-patterns in the top and bottom interfaces have equal moir\'{e}-wavelengths and their relative misorientation (see Eqs.2,3, SI \cite{supp}) is one of the commensurate angles satisfying
\begin{equation}
    \left|\Phi_{12} - \Phi_{23}\right| = \theta_{\text{r}} = 2\tan^{-1}(p/\sqrt{3}q)
    \label{eqn:commcondition}
\end{equation}
then $(m,n)=(1,1)$ becomes the dominant SM wavelength and is written in terms of the two coprime $q,p$ as
$ \lambda_{\text{SM}} =
\frac{\sqrt{3q^{2}+p^{2}}}{2p}\,\lambda_{0}$.
All the points in the parameter-space where an exact periodicity exists lie along the line $\theta_{12} = -\theta_{32}$ as in Fig.\ref{fig:systemtri}(b2) and (c2) for two different type of trilayer SMS. This SM wavelength is identically equal to the magnitude of direct-space PV in the SM structure for odd $q$ and $p=1$ but might differ for other cases \cite{Shallcross2010}. For $q=2n+1$ with $n\in \mathbb{N}$ and $p=1$, one has $\lambda_{\text{SM}} = \sqrt{3n^{2}+3n+1}\,\lambda_{0}$ the sequence for different values of $n$ is $\sqrt{7},\sqrt{19}, \sqrt{37},\dots$ which we refer to as a L\"{o}schian sequence following \cite{loschbook,Kitchin2009,Aggarwal2024}. \textit{E.g.}, for $\sqrt{7}$, $7$ copies of the SM BZ fit within one BZ of either the top or bottom moir\'{e} as in Fig.~\ref{fig:systemtri}(d). The outer boundary of these $7$ hexagons that overlap with one hexagon form a particular iteration of an iterative fractal both in real- and RS where a unique fractal generator (see Fig.~\ref{fig:systemtri}(d)) is then applied to an initiator \cite{Arlinghaus1985,Arlinghaus1989} which is the moir\'{e} BZ of either the top or bottom interface.

\begin{figure*}
	\centering
	\includegraphics[width=\linewidth]{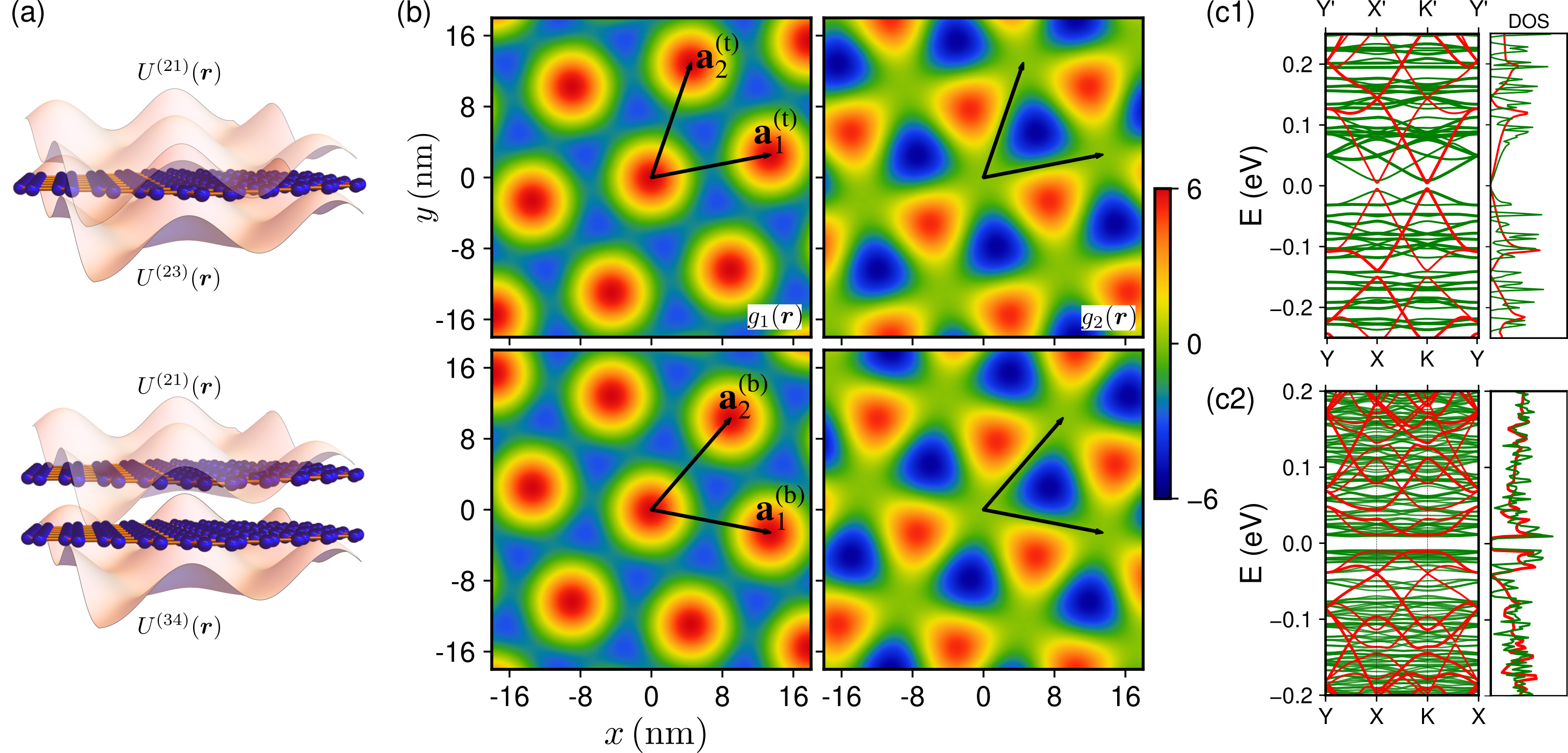}
	\caption{
 (a) Structures leading to the effective Hamiltonian of an AAA-stacked hBN-encapsulated single-layer graphene and an AAA-stacked hBN-encapsulated Bernal-stacked bilayer graphene. The layers are indexed in both configurations using $ \ell=1,2,3 $ from the top to bottom. The potentials $U^{(21)}$, $U^{(23)}$, and $U^{(34)}$ show a visual representation of the Eqs.\ref{eqn:Hambn2} and \ref{eqn:Hamblg2}. The moir\'{e} fractal corresponding to $(q,p) = (3,1)$ for both the cases $viz.$ hBN-encapsulated SLG and hBN-encapsulated BLG is identical to Fig.\ref{fig:systemtri}(d) where the two solid lines in maroon and dark blue represent the BZ of both the hBN layers and the black solid line represents the seven copies of the BZ of the supermoir\'{e} configuration.
 (b) The pseudocolor spatial distribution of the functions $g_{1}(\bmr)$ and $g_{2}(\bmr)$ in both the top and bottom moir\'{e} interfaces. In the top row, the $g_{1}(\bmr)$ and $g_{2}(\bmr)$ of the top moir\'{e} interface are rotated anticlockwise, while the $g_{1}(\bmr)$ and $g_{2}(\bmr)$ of the bottom moir\'{e} interface are rotated clockwise. The PV in both the moir\'{e} interfaces are shown by black arrows. (c1) and (c2) show the bandstructures and density of states that correspond to hBN-Graphene-hBN for $(q,p) = (3,1)$ and hBN-BLG-hBN also for $(q,p) = (3,1)$, respectively. In both cases, the number of bands within the gap of the lowest-four bands (with valley degeneracy) remains identically $ 4\left(e^{\ln(n_{c})/L_{N}} - 1\right) $, where $L_{N}$ is a L\"{o}schian number \cite{Marshall1975,Arlinghaus1985,Arlinghaus1989}.}
	\label{fig:systemtrihbn}
\end{figure*}

We demonstrate this in a trilayer supermoir\'{e} system where a G-layer \cite{Wang2019, Viti2020, Ezzi2024} or a BLG \cite{Lujun2019, Zihao2019, Finney2019, Sun2021} is sandwiched between two hBN layers. For the latter, quad-layer system (FIG.~\ref{fig:systemtrihbn}(c)), since the Bernal-stacked BLG is rotated as a whole, effectively, this is also taken as a trilayer SMS. 

For smaller twist angles 
of the hBN layers \emph{wrt} the middle graphene layer and for wavevectors $k \ll K $, the long-wavelength description is sufficient. In this low-energy regime, the degrees of freedom due to the two hBN layers integrate out, and one writes an effective $2\times2$ Hamiltonian for single or Bernal stacked BLG in the two-sublattice basis in an effective potential (see Sec-II, SI \cite{supp}), such that,
\bea
	H^{(1)}_{\text{eff}}(\bmr) & = &  -i\hbar v_{F}\bm{\sigma}_{\xi}\cdot\bm{\nabla} + \eta \big[ U^{(21)}_{\xi}(\bmr) + U^{(23)}_{\xi}(\bmr) \big]
	\label{eqn:Hambn2} \\
	H^{(2)}_{\text{eff}}(\bmr) & = &  
	\begin{bmatrix}
		h^{(2)}_{\text{G},\xi} + U_{21}(\bmr) & \tilde{V}^{(23)}_{\xi} \\
		\tilde{V}^{(23),\dagger}_{\xi} & h^{(3)}_{\text{G},\xi} + U_{34}(\bmr)
	\end{bmatrix}
	\label{eqn:Hamblg2}
\eea 
where $v_{F} \approx 10^{6}\,\unit{\meter\second^{-1}}$ is the Fermi velocity in graphene, $U^{(\ell \ell')}(\bmr)$ is the periodic effective potential due to the hBN layer-$\ell'$ on the graphene layer-$\ell$ includes interlayer hopping terms (details in Eqs.10,11, SI \cite{supp} and refs. \cite{Yankowitz2012,Wallbank2013,Jose2014Aug,Jose2014Sep}) given as
\begin{equation}
    U^{(\ell \ell')}(\bmr) =
    V^{(\ell \ell'),\dagger}(\bmr)\,h^{(\ell'),-1}_{\text{hBN}}(\bmr)\,V^{(\ell \ell')}(\bmr)
    \label{eqn:moir\'{e}pot}
\end{equation}
We shall assume that the above description is valid throughout this work. Eq.\ref{eqn:Hambn2} and Eq.\ref{eqn:Hamblg2} can equivalently be considered as a massless Dirac fermion in graphene layer(s) to two superlattice periodic external potentials having different moir\'{e} periodicities given by \ref{eqn:moir\'{e}pot}) and schematically depicted in Fig. \ref{fig:systemtrihbn}(a). 
The moir\'{e} periodicities come from $g_{1}(\bmr)$ and $g_{2}(\bmr)$ (Eqs.26,27, SI \cite{supp}), plotted in Fig.\ref{fig:systemtrihbn}(b). Bloch's theorem can now be applied for a common periodicity in commensurate SMS, giving the state $\ket{n\bmk,\xi}$ at each valley $\xi$.

Following Eq.\ref{eqn:commcondition}, for a commensurate-angle $\theta_{r} = \ang{21.79}$ at $ (q,p)=(3,1)$ between the two moir\'{e}-interfaces, the twist-angle of the top G-hBN interface is $\theta_{12} \sim -\ang{0.1} $ and the twist-angle of the bottom G-hBN interface becomes $\theta_{32} \sim \ang{0.1} $ with a supermoir\'{e} wavelength of about $\lambda_{\text{SM}} \sim 36.13\,\unit{\nano\meter}$. The orientations of the moir\'{e}-pattern by $\Phi_{12} \sim \ang{10.89}$ of the top interface and by $\Phi_{32} \sim -\ang{10.89}$ of the bottom interface is due to the small twists of the top and bottom hBN layers for $(q,p) = (3,1)$. For this configuration, the ratio of the area of the moir\'{e} BZ of either interface to the super minizone corresponding to the SM length scale are identically $7$ as in Fig.\ref{fig:systemtri}(d). The boundaries of the super minizone overlap with both the BZ of two moir\'{e} interfaces. This boundary can be generated by applying the generator corresponding to $(q,p) = (3,1)$ to either of the red hexagons as in the subfigure of Fig.\ref{fig:systemtri}(d). With this common periodicity in SM wavelength, we show the bandstructure of hBN-graphene-hBN in Fig.\ref{fig:systemtrihbn}(c1). The red bandstructure corresponds to a single G-hBN moir\'{e} interface (top) for strained parameters in Table-\ref{table:ghbnparams}; the green bandstructures to strained hBN-graphene-hBN parameters. The BS for hBN-BLG-hBN is shown in Fig.\ref{fig:systemtrihbn}(c2). Due to the presence of two hBN layers on the top and bottom, the parameters $w_{0}$ and $\tilde{w}_{3}$ add twice of strength \textit{viz-z-viz} only one moir\'{e}-interface. The high-symmetry path $Y-X-K-Y$ shows the bandstructure for the right-valley ($\xi=1$) of graphene, while $Y'-X'-K'-Y'$ shows the bandstructure for the left-valley ($\xi=-1$) for both the cases. This implies that the insertion of the new bands remains consistent with the different set of parameters. Within the bandwidth of the lowest valence band in the graphene-hBN system, there are $14$ bands of both the valleys of the hBN-graphene-hBN system. In general, this total number of bands within the bandwidth is related to $D_{f}$ of the fractals corresponding to a given $(q,p)$ as $g_{s}g_{v}e^{2\ln{(D_{f})}/n_{c}}$ \cite{Aggarwal2024}, where $n_{c}$ is the number of sides in the generator and $g_{s}$ and $g_{v}$ are the spin and valley degeneracies.

\begin{figure*}
    \centering
    \includegraphics[width=\linewidth]{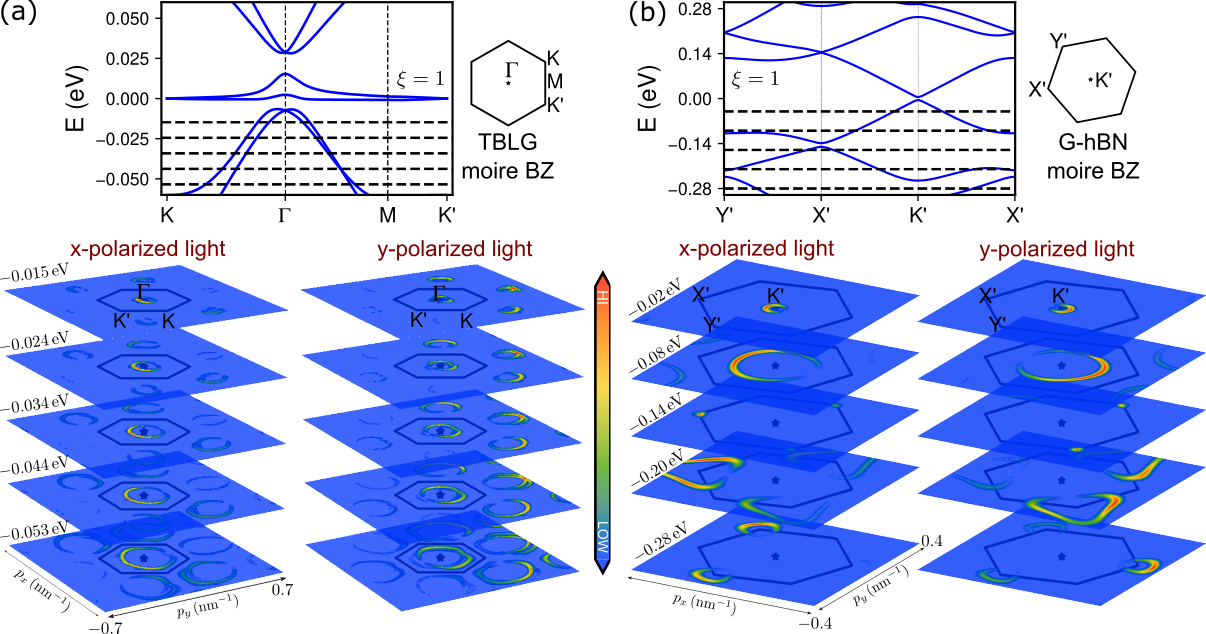}
    \caption{
Isoenergy ARPES contour momentum distributions for (a) relaxed magic-angle twisted bilayer graphene and (b) graphene-hBN interface with a strained graphene layer closer to hBN at different photoelectron energy with $x$- and $y$-polarized light states. The colour bar shows the ARPES intensity obtained with $x$- and $y$- polarization states of light in $ I(\bm{p}, E) $ with a Gaussian smearing factor 0.02. The top rows show the right-valley bandstructure along the high-symmetry path, which is also indicated in the BZ adjacent to the bandstructure plots. The dashed line below the $E_{f}$ in the valence band region marks the energy levels at which the ARPES intensity calculation is performed. These lines are also indicative of the ARPES contour features, \textit{e.g.}, the dashed lines in (a) crossing the bands only near the $\Gamma$-point and therefore, the ARPES contribution comes from the region around the $\Gamma$-point. Similarly, the ARPES contribution in the case of G-hBN moir\'{e} interface at different levels of energy shift from the $K'$-point to the corners of the BZ. The colourmaps are drawn using a power norm with exponent $1.1$.}
    \label{fig:arpes_oneinterface}
\end{figure*}

Angle-resolved photoemission spectroscopy (ARPES) suitably designed for moir\'{e} superlattice length case often dubbed as nano-ARPES \cite{Avila_2013_1, Avila_2013_2, Diaz2015, LIM_2024}, can provide precise information on the newly-inserted bands in hBN-graphene(BLG)-hBN supermoir\'{e} structures consistent with $D_f$ from Fig. \ref{fig:systemtri}(d) by probing the momentum-dependence of the one-particle electronic Green's function and hence their count within the bandwidth of the lowest valence band of one graphene-hBN interface. For a periodic lattice with different sublattices $ \alpha $, the ARPES intensity of an emitted photoelectron as a function of its 3-momentum $ \bm{p} $ and energy $ E $ is
\begin{equation}
	I(\bm{p},E) \propto \sum_{\xi} \sum_{n,\bmk} \abs{
    \bm{A} \cdot \mel{\bm{p}}{\hat{\bm{v}}}{n\bmk,\xi}}^{2}
	\delta\left(E-\epsilon^{\xi}_{n\bmk}\right)
 \label{eqn:arpesform1}
\end{equation}
where $ \bm{A} $ is the magnetic vector potential of the incident electromagnetic wave, $\hat{\bm{v}}$ is the velocity operator, $ \bm{p}_{\parallel} $ is the 3-momentum of photoelectron near the $ \bm{K}^{\xi} $-point. The $\ket{n\bmk,\xi}$ is the Bloch state of the underlying system. The incoming light selects a specific valley $\xi$ and a reciprocal lattice vector to map the Bloch wave vector $\bmk$ into the first BZ. In Fig.\ref{fig:arpes_oneinterface} (a) and (b), we showed respectively the momentum distribution functions measured at different energies for the relaxed TBLG, where we included the effect of the strain fields on the ARPES intensity as compared to earlier studies \cite{Zhu2021} (Eq.28 in SI \cite{supp} and subsequent equations), and the top moir\'{e} interface in hBN-encapsulated single-layer graphene using the Hamiltonian in Eq.\ref{eqn:Hambn2} with only the moir\'{e}-potential of the top interface $U^{(12)}(\bmr)$.
We ignore the photon-energy dependence of the ARPES intensity by using the $p_{z} = 50\,\unit{\nano\meter}^{-1}$ (corresponding to $100\,\unit{\milli\electronvolt}$ photon) while considering energies up to $\sim 150\,\unit{\milli\electronvolt}$ specifically for TBLG. In both figures, the energies at which the momentum distributions are calculated are indicated in the bandstructure plots in the top panels of Fig.\ref{fig:arpes_oneinterface}(a) and (b) by black dashed lines. The different hopping parameters partially account for strain and corrugation effects, and the ARPES signals significantly depend on their ratio. In addition, the shape of the constant-energy ARPES contours depends on the state of polarization of the incident light. This contribution rotates by $\pi$ when a $y$-polarization is allowed compared to $x$-polarized light. $U^{(12)}(\bmr)$ leads to the appearance of secondary Dirac cones \cite{Yankowitz2012, MKinder2012,Wallbank2013} in the energy spectrum, and the constant-energy ARPES contours in Fig.\ref{fig:arpes_oneinterface}(b) at energies near the Fermi level ($E_{f}$) show a large contribution from these additional Dirac points near the high-symmetry $X'$-point for energies $-0.14\,\unit{\electronvolt}$ and $-0.2\,\unit{\electronvolt}$. In particular, the ARPES momentum-distribution functions at lower energies are similar to the circular energy surface of monolayer graphene, whereas far from the $E_{f}$ the contribution comes largely from the corners of the moir\'{e} BZ making the constant-energy contour maps anisotropic functions of the momentum direction.

\begin{figure*}
    \centering
    \includegraphics[width=\linewidth]{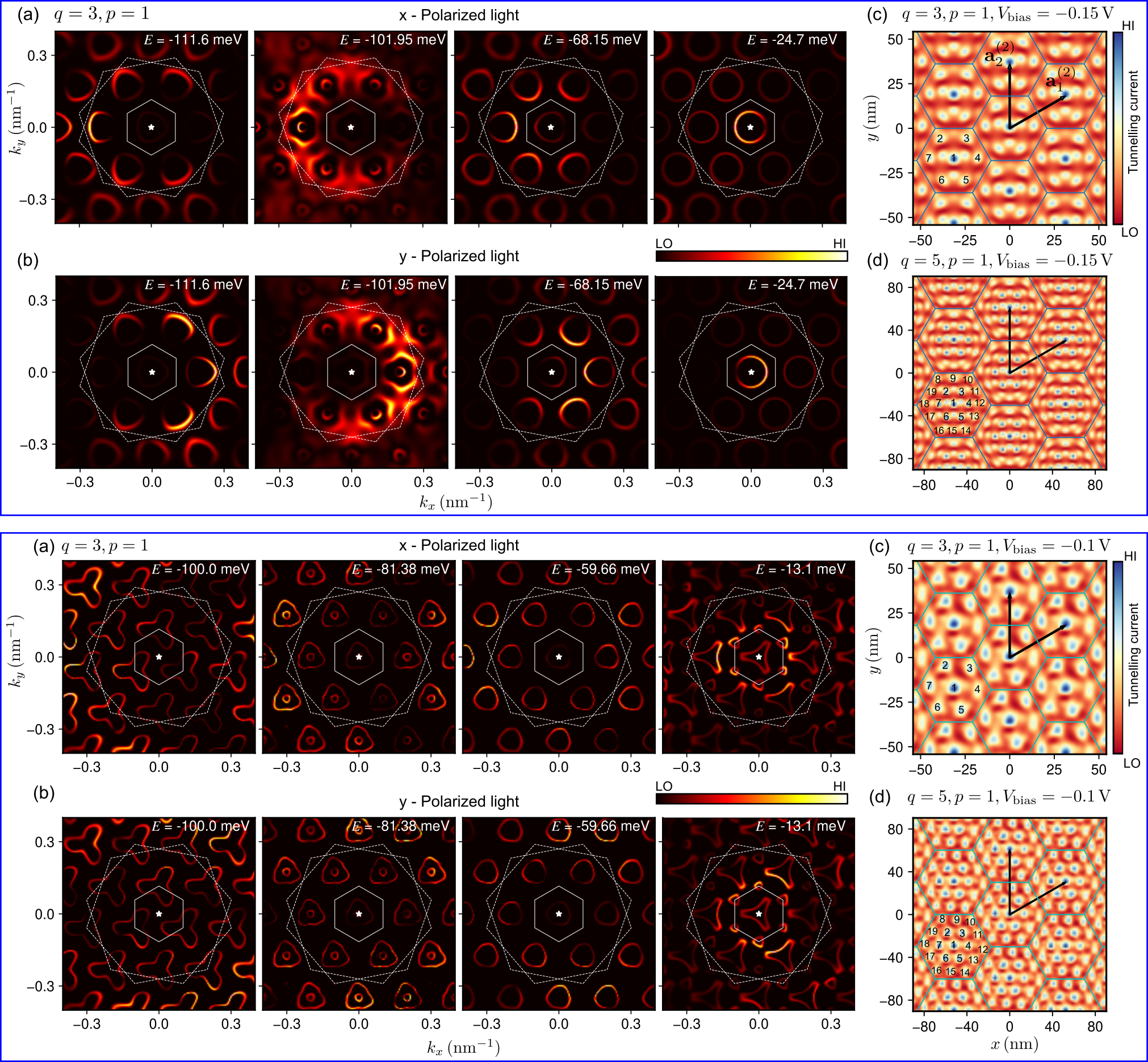}
    \caption{
Top panel: (a) and (b) show the isoenergetic ARPES contour momentum distributions for hBN-encapsulated single-layer graphene for different photoelectron energies with $x$- and $y$-polarized light states for $(q,p)=(3,1)$. The white dashed lines show the BZs of the moir\'{e}-patterns in two interfaces, and the solid white line shows the BZ of the resulting commensurate supermoir\'{e} structure. The star at the centres represents the high-symmetry $\Gamma$-point. The ARPES intensities are calculated using a Gaussian smearing factor of $ 0.005 $. (c),(d) show the tunneling current ($I_{t}$) in hBN-encapsulated single-layer graphene corresponding to two different supermoir\'{e} commensurations $(q,p)=(3,1)$ and $(q,p)=(5,1)$ respectively. The bias voltage in the current calculation is set to be $V_{\text{bias}} = -0.15\,\unit{\volt}$. The two vectors $\bm{a}^{(2)}_{1},\bm{a}^{(2)}_{2}$ show the primitive lattice vectors in the supermoir\'{e} structures and the number of peaks that resonates with the predicted precounted number of bands count is marked in black. Bottom panel: A similar analysis is performed for the hBN-BLG-hBN. The only difference is the bias voltage $V_{\text{bias}} = -0.1\,\unit{\volt}$.}
    \label{fig:hbnghbn_stm}
\end{figure*}

The constant-energy ARPES contours for SMS structures: hBN-G-hBN and hBN-BLG-hBN for the $x$-polarization state of the incident radiation, are shown respectively in the top and bottom panel of Fig.\ref{fig:hbnghbn_stm}(a) with corresponding energy levels mentioned in the right-corner of each plot corresponding to bandstructure plot in Fig.\ref{fig:systemtrihbn}(c) using the continuum Hamiltonian in Eq.\ref{eqn:Hambn2}, and the bandstructure plot Fig.\ref{fig:systemtrihbn}(d) using the continuum Hamiltonian in Eq.\ref{eqn:Hamblg2}. For both cases, the $\Gamma$-point in the centre is shown by a star, and the two bigger rotated dashed-line hexagons identify the moir\'{e} BZs of the two moir\'{e} interfaces in hBN-G/BLG-hBN, and the inner solid-line hexagon marks the supermini or supermoir\'{e} BZ. For hBN-G-hBN in the first three contour plots, within the dashed-line hexagons but outside the solid-line hexagon, the ARPES contribution stems from six points that are the secondary Dirac points of the supermoir\'{e} system. Similarly, for hBN-BLG-hBN, within the dashed-lines hexagons but outside the solid-line hexagon, the ARPES contribution stems from six closed polygons with the centres of these polygons being the secondary Dirac points of the supermoir\'{e} system. The fundamental difference between the parabolic bands in hBN-BLG-hBN and the linear bands in hBN-G-hBN results in starkly different closed polygons around the Dirac points as shown in (a) and (b) of the top and bottom panels of Fig.\ref{fig:hbnghbn_stm}.
 
For either SMS, the intensity distributed around each of the Dirac points is highly anisotropic, reflecting $C_{3}$-symmetry of the wave functions. For the hBN-G-hBN SMS, the intensity primarily originates from the left Dirac points, whereas the contribution from the central primary Dirac point and one of the six secondary Dirac points on the right remains low. In contrast, near the $E_{f}$ in the bandstructure, the primary Dirac point dominates and also remains anisotropic. The primary and secondary Dirac points collectively count to $7$ within the dashed-line moir\'{e} BZs of both the interfaces, confirming the iterative fractal nature of such trilateral SMS when the two moir\'{e} interfaces are commensurate. For the hBN-BLG-hBN SMS, the contour plot at $E = -13.10\,\unit{\milli\electronvolt}$ shows not only the intensity contribution from the Dirac points but also the contribution from the corners of the BZ. The polygons at three equivalent corners are smaller than the polygons of the remaining equivalent three corners. The brightest intensity is distributed near the boundary of the supermoir\'{e} BZ. Again, the primary and secondary Dirac points collectively count to $7$ within the dashed-line moir\'{e} BZs of both the interfaces, which evidently confirms the addition of a \emph{countable} number of bands in this trilayer SMS when the two moir\'{e} interfaces are commensurate. This underscores the possibility of the experimental visualization of the extra \emph{pre}-countable number of quantum states in such SMS, demonstrating their fractal nature. The $\pi$-phase change for a $y$ to \emph{x}-polarized light in the intensity contribution is clearly visible in Fig.\ref{fig:hbnghbn_stm}(b) top panel.
 
\begin{table}
	\centering
	\caption{The parameters in the effective Hamiltonian of Eq.~\ref{eqn:moir\'{e}pot} in the presence of the lattice relaxation. All numbers are in \unit{\milli\electronvolt} \cite{Jose2014Sep}.}
	\begin{ruledtabular}
		\begin{tabular}{ccccccccc}
			& $w_{0}$ & $\tilde{w}_{3}$ & $u_{0}$ & $\tilde{u}_{0}$
			& $u_{3}$ & $\tilde{u}_{3}$ & $u_{\perp}$ & $\tilde{u}_{\perp}$ \\
			\hline
			unstrained & 12 & 0 & 1 & 4.4 & 1.8 & -7.6 & 2.1 & -8.8 \\
			strained & 3.9 & 5.3 & 2 & 5.2 & -0.06 & -5.9 & 21 & -42 \\
		\end{tabular}
	\end{ruledtabular}
	\label{table:ghbnparams}
\end{table}

 The differential conductivity $\dd{I_{t}}/\dd{V}=(2e\zeta / \hbar^{2})\,\rho(\bmr,E)$, where the tunneling current $I_{t} \propto \sum_{E=E_{F} - eV_{\text{bias}}}^{E_{F}} \rho(\bmr, E)$, $\rho(\bmr,E)$ is the local density of states (LDOS), and $e$ is the electron charge, can be probed directly using a scanning tunneling microscope (STM) \cite{Coe2024} with $\zeta$ containing the details of the STM tip, is another quantity that can reveal the details of the surface topography of such SMS and identify the fractality in appropriate proportion in real space, complementing the confirmation of ARPES. The LDOS, 
\bea
    \rho(\bmr, E) & = &  -\frac{1}{\pi} \Tr{\Im{G(\bmr,\bmr;E)}} \nonumber \\
    \text{with}, G^{\alpha\alpha'}(\bmr, \bmr' ; E)
    & =&  \sum_{n,\bmk} \frac{\braket{\bmr}{n,\bmk,\alpha}\braket{n,\bmk,\alpha'}{\bmr'}}{E - E_{n\bmk} + i\eta^{+}}
    \label{eqn:green_fun}
\eea
is the local single-particle retarded on-site Green's function. In evaluating spatial maps corresponding to $\dd{I_{t}}/\dd{V}$, we drop the spin-index $\sigma$ and set right valley $\xi = 1$.

$I_{t}$ for hBN-G-hBN and hBN-BLG-hBN for two structural configurations $(q,p) = (3,1)$ and $(q,p) = (5,1)$ are shown in (c) and (d) respectively in the top and bottom panel of Fig.\ref{fig:hbnghbn_stm} at bias voltage $V_{\text{bias}} = -0.15\,\unit{\volt}$ and $V_{\text{bias}} = -0.1\,\unit{\volt}$ for the right-valley $\xi=1$. The light blue lines identify the real-space lattice of the supermoir\'{e} lattice for the two commensurate moir\'{e}-interfaces. The two black arrows are the real-space primitive lattice vectors $\bm{a}^{(2)}_{i}$ for $i=1,2$ of the SMS. Each hexagon shows a copy of the primitive unit cell with its centres contributing maximally to $I_t$. This periodic distribution of $I_{t}$ in each cell is due to the Bloch periodicity of the local density of state $\rho(\bmr,E)$ with the PV of the lattice. The different number of enclosed Wigner-Seitz (WS) cells of each moir\'{e}-interface within the WS cell of the SMS at $(q,p)=(5,1)$ \textit{viz-a-viz} $(q,p)=(3,1)$ can be understood by comparing (d) to (c). The number of maxima in $I_{t}$ within a cell is representative of the newly inserted bands within the bandgap as controlled by the $(q,p)$-integers in the SM structure. These peaks are enumerated in the contour maps. \textit{Vs.} the central maxima, the other peaks in the neighbourhood are elongated and show anisotropy of $I_{t}$ in real space. 

The sharp features in the contour plots are highly sensitive to the parameter $\eta^{+}$ in the spatially resolved single-particle Green's function in Eq.~\ref{eqn:green_fun}. There is no universally robust method for determining $\eta^{+}$ from the available eigenvalue data $E_{n\bmk}$. Typically, $\eta^{+}$ is chosen to enhance local maxima while simultaneously suppressing noise. Within the WS cell, the local maxima, indicated by the blue regions in the contour plots, form a valley-like structure, with the valleys occurring at the centres of each moiré-interface cell. This distribution of $I_{t}$ provides a robust way to visualize the additional pre-counted states introduced by the presence of an additional moiré potential, rotated to a commensurate angle, on top of the existing moiré potential in both tri- and quad-layer cases.

To conclude, we have predicted a specific type of iterative fractal in two well-known experimentally available SMS: hBN-encapsulated mono- and Bernal-stacked BLG and pointed out robust experimental signatures accessible to ARPES and STM that can extract the precise number of inserted bands in the low-energy regime of such quantum iterative fractals in twisted vdW heterostructures, thus confirming their $D_f$ similar to Ref.\cite{LaiAndrei2024} where a moir\'{e} phase diagram is unveiled by carefully analyzing the STM of TBLG on hBN. Our methodology can be applied to any SMS arising from homo- and hetero-bilayer moir\'{e} materials, provided they possess a gap in their spectrum. Particularly in a homo-bilayer system, it can be done with suitable gating. Iterative fractality can be exploited to control band insertions in the energy spectrum, thereby further flattening the moir\'{e} bands in such SMS, thus yielding a controlled way to further localize the electronic states and opening the possibility of engineered moir\'{e} materials. Our analysis provides a pathway to understanding the hitherto-unexplored bandstructure-related complexity in a generic super moir\'{e} system in an experimentally accessible region of parameter space while simultaneously opening a technological front towards designer quantum materials.

\begin{acknowledgments}
SG is supported by MTR/2021/000513 funded by SERB, DST, Govt. of India. DA is supported by a UGC fellowship and SPARC project P2117 by MoE, Govt. of India.
\end{acknowledgments}

\bibliography{reference}
\end{document}